\documentclass[runningheads]{llncs}
\usepackage[utf8]{inputenc}
\usepackage[pdftex]{graphicx}
\usepackage[cmex10]{amsmath}

\usepackage{floatrow}
\DeclareFloatFont{tiny}{\scriptsize}
\usepackage{subcaption}
\usepackage[hyphens]{url}
\usepackage[table]{xcolor}
\usepackage{xspace}
\PassOptionsToPackage{hyphens}{url}\usepackage{url}
\usepackage{tikz}
\usepackage{multirow}
\usepackage{pgfplots}
\usetikzlibrary{pgfplots.groupplots}
\usepackage{letltxmacro}

\usepackage{cite} 
\pgfplotsset{compat=newest}
\usepackage{ifthenx}
\usepackage{collcell}
\usepackage{nicefrac}
\usepackage{amsmath}
\usepackage{amsfonts}
\usepackage{filecontents}
\usepackage{eucal}
\usepackage{listings}
\usepackage{lipsum}
\usepackage{todonotes}
\usepackage{makecell}
\usepackage{paralist}
\usepackage{flushend}
\usepackage{csvenhanced}
\usepackage{siunitx,booktabs}

\LetLtxMacro{\oldtodo}{\todo}
\renewcommand{\todo}[2][]{\tikzexternaldisable\oldtodo[fancyline,size=\footnotesize,#1]{#2}\tikzexternalenable}
\renewcommand{\todo}[1]{\tikzexternaldisable\oldtodo[fancyline,size=\footnotesize]{#1}\tikzexternalenable}

\usepackage{algorithmic}
\usepackage[ruled]{algorithm2e}

\usepackage{color}
\usepackage{paralist}
\lstdefinestyle{mystyle}{
    commentstyle=\color{green!40!black},
    basicstyle=\scriptsize\ttfamily,
    breakatwhitespace=false,
    breaklines=true,
    captionpos=b,
    keepspaces=true,
    numbers=left,
    numbersep=5pt,
    showspaces=false,
    showstringspaces=false,
    showtabs=false,
    tabsize=2,
    xleftmargin=15pt,
    numberbychapter=false
}
\newcommand{\etal}{et~al.} 
\newcommand{\ie}{\textit{i.e.}, } 
\newcommand{\eg}{e.g., } 
\newcommand{\cf}{c.f., } 
\mathchardef\mhyphen="2D

\newcommand{\xor}{\oplus}

\begin{document}

\title{Rowhammer.js: A Remote Software-Induced Fault Attack in JavaScript}

\author{Daniel Gruss \and Clémentine Maurice$^{\dagger}$ \and Stefan Mangard}
\authorrunning{Daniel Gruss \and Clémentine Maurice \and Stefan Mangard}
\institute{Graz University of Technology, Austria}

\maketitle

\begin{abstract}
A fundamental assumption in software security is that a
memory location can only be modified by processes that may write to this
memory location. However, a recent study
has shown that parasitic effects in DRAM can change the content of a memory cell
without accessing it, but by accessing other memory locations in a high frequency.
This so-called Rowhammer bug occurs in
most of today's memory modules and has fatal consequences for the
security of all affected systems, \eg privilege escalation attacks.

All studies and attacks related to Rowhammer so far rely on the
availability of a cache flush instruction in order to cause
accesses to DRAM modules at a sufficiently high frequency. We overcome this limitation by defeating complex
cache replacement policies. We show that caches can be forced into
fast cache eviction to trigger the Rowhammer bug with only regular
memory accesses. This allows to trigger the Rowhammer bug in highly
restricted and even scripting environments.

We demonstrate a fully automated attack that requires nothing but
a website with JavaScript to trigger faults on remote hardware.
Thereby we can gain unrestricted access to systems of website visitors.
We show that the attack works on off-the-shelf systems.
Existing countermeasures fail to protect against this new
Rowhammer attack.
\end{abstract}

\bgroup
\let\thefootnote\relax\footnotetext{This paper has been accepted at DIMVA 2016 (\url{dimva2016.mondragon.edu/en}). The final publication is available at link.springer.com (\url{http://link.springer.com/}).}
\let\thefootnote\relax\footnotetext{$^\dagger$ Part of the work was done while author was affiliated to Technicolor and Eurecom.}
\egroup

\section{Introduction}
Hardware-fault attacks have been a security threat since the first
attacks in 1997 by Boneh~\etal\cite{DBLP:conf/eurocrypt/BonehDL97} and
Biham~\etal\cite{DBLP:conf/crypto/BihamS97}. Fault attacks typically
require physical access to the device to expose it to physical
conditions which are outside the specification. This includes high or
low temperature, radiation, as well as laser on dismantled microchips.
However, software-induced hardware faults are also possible, if the
device can be brought to the border or out of the specified operation
conditions using software.
Kim~\etal\cite{Kim2014ISCA} showed that frequently accessing specific
memory locations can cause random bit flips in DRAM chips. $85\%$ of the
DDR3 modules they examined are vulnerable. The number of bit flips
varies from one module to another, \ie some modules can be more
vulnerable than others. More recently, DDR4 modules have been found to
be vulnerable as well~\cite{Pessl2015}.
Bit flips can be triggered by software by flushing a memory location
from the cache and reloading it. Seaborn~\cite{SeabornBlackhat2015}
demonstrated that an attacker can exploit such bit flips for privilege
escalation. These exploits are written in native code and use special
instructions to flush data from the cache.

We show that it is possible to trigger hardware faults by performing
fast cache eviction on all architectures, if the DRAM modules are
vulnerable. Compared to previous work, we do not use any specific
instruction, but only regular memory accesses to evict data from the cache.
The attack technique is thus generic and can be applied to any
architecture, programming language and runtime environment that allows
producing a fast stream of memory accesses. Therefore, proposed
countermeasures such as removing the \texttt{clflush} instruction cannot
prevent attacks. Even more severe, we show that on vulnerable modules, we
can also perform remote JavaScript-based Rowhammer attacks.

Since an attack through a
website can be performed on millions of victim machines simultaneously
and stealthily, it poses an enormous security threat.
Rowhammer.js is independent of the instruction set of the CPU. It is the
first remote software-induced hardware-fault attack. As a proof of
concept, we implemented a JavaScript version that as of today runs in
all recent versions of Firefox and Google Chrome.

For a Rowhammer attack in JavaScript we perform the following steps:
\begin{compactenum}
\item Find 2 addresses in different rows
\item Evict and reload the 2 addresses in a high frequency
\item Search for an exploitable bit flip
\item Exploit the bit flip (\eg manipulate page tables, remote code execution)
\end{compactenum}
Steps 3 and 4 have already been solved in previous work~\cite{SeabornBlackhat2015}, but step 1 and 2 remain open challenges.

The challenge in step 1 is to retrieve information on the physical
addresses from JavaScript. It is strictly sandboxed and provides
no possibility to retrieve virtual or physical addresses. To tackle this
challenge, we determine parts of the physical addresses using large
arrays that are allocated by operating systems on large pages. We thus
do not exploit any weaknesses in JavaScript or the browser, but only
OS-level optimizations.

The challenge in step 2 is to find fast cache eviction strategies to replace the \texttt{clflush} instruction.
On older CPUs, simply accessing $n+1$ addresses is sufficient to evict
lines for an $n$-way cache~\cite{Liu2015,Oren2015CCS}.
On Intel CPUs produced in the last 4 years, \ie post Sandy Bridge, the
replacement policy has changed and is undocumented. Consequently,
known eviction strategies have a low eviction rate or a high
execution time, which is not suitable for Rowhammer attacks.
To tackle
this challenge, we present a novel generic method for finding cache eviction strategies that achieve the best performance in both
timing and eviction rate by comprehensively exploring the parameter space.
We present the best eviction strategies so far, outperforming previous ones on all recent Intel architectures.
Based on this method, we build a two-phase online attack for remote systems with unknown hardware configuration.

\begin{table}[t]
\caption{Experimental setups.}\label{tbl:machines}
\resizebox{\hsize}{!}{
\begin{tabular}{llll}
\toprule
Platform & CPU & Architecture & RAM \\
\midrule
Lenovo T420 & i5-2540M &  Sandy Bridge & Corsair DDR3-1333 8 GB \\
& & & and Samsung DDR3-1600 4 GB ($2\times$) \\
Lenovo x230 & i5-3320M & Ivy Bridge & Samsung DDR3-1600 4 GB ($2\times$) \\
Asus H97-Pro & i7-4790 & Haswell & Kingston DDR3-1600 8 GB \\
ASRock Z170 ITX & i7-6700K & Skylake & G.Skill DDR4-3200 8 GB ($2\times$) \\
& & & and Crucial DDR4-2133 8 GB ($2\times$) \\
\bottomrule
\end{tabular}
}
\end{table}

We compare the different implementations of the Rowhammer attacks on a
fixed set of configurations (see Table~\ref{tbl:machines}), some
vulnerable in default settings, others at decreased refresh rates.

As of today, software countermeasures against Rowhammer native code
attacks only target specific exploits, and, as we show, do not protect
sufficiently against attacks from JavaScript.
Hardware countermeasures are harder to deploy, since they do not affect
legacy hardware including recent vulnerable DDR4 modules.
BIOS updates can be used to solve the problem on commodity systems,
however it is only a practical solution for very advanced users.

Summarizing, our key contributions are:
\begin{itemize}
\item We provide the first comprehensive exploration of the cache
eviction parameter space on all recent Intel CPUs. This also benefits
broader domains, \eg cache attacks, cache-oblivious algorithms, cache
replacement policies.
\item We build a native code implementation of the Rowhammer attack that
only uses memory accesses. The attack is successful on Sandy Bridge, Ivy
Bridge, Haswell and Skylake, in various DDR3 and DDR4 configurations.
\item We build a pure JavaScript Rowhammer implementation, showing that
an attacker can trigger Rowhammer bit flips remotely, through a web browser.
\end{itemize}

The remainder of this paper is organized as follows.
In Section~\ref{sec:background}, we provide background information on
DRAM, the Rowhammer bug, CPU caches, and cache attacks.
In Section~\ref{sec:opt_cache_evict}, we describe a two-phase automated
attack to trigger bit flips on unknown systems.
In Section~\ref{sec:rowhammer_without_flush}, we demonstrate the
Rowhammer bug without \texttt{clflush} in native code and in
JavaScript.
In Section~\ref{sec:discussion}, we provide a discussion of our
proof-of-concept exploit, limitations, and countermeasures.
Finally, we discuss future work in Section~\ref{sec:future} and provide
conclusions in Section~\ref{sec:conclusions}.

\section{Background}\label{sec:background}

\subsection{DRAM}
Modern memory systems have multiple \textit{channels} of DRAM memory connected to the memory controller. 
A channel consists of multiple \textit{Dual Inline Memory Modules} (\textit{DIMMs}), that are the physical modules on the motherboard. 
Each DIMM has one or two \textit{ranks}, that are the sides of the physical module.
Each rank is a collection of \textit{chips}, that are further composed of \textit{banks}. Accesses to different banks can be served concurrently.
Each bank is an array of capacitor cells that are either in a charged or discharged state, representing a binary data value. 
The bank is represented as a collection of rows, typically $2^{14}$ to $2^{17}$.

The charge from the cells is read into a \textit{row buffer} on request and written back to the cells as soon as another row is requested.
Thus, access to the DRAM is done in three steps: 
\begin{inparaenum}
	\item opening a row,
	\item accessing the data in the row buffer,
	\item closing the row before opening a new row, writing data back to the cells.
\end{inparaenum}

DRAM is volatile memory and discharges over time. The \textit{refresh interval} defines when the cell charge is read and restored to sustain the value. DDR3 and DDR4 specifications require refreshing all rows at least once within 64ms~\cite{Kim2014ISCA,Aichinger2015}.

The selection of channel, rank, bank and row is based on physical address bits. The mapping for Intel CPUs has recently been reverse engineered~\cite{Seaborn2015DRAMmap,Pessl2015}.

\subsection{The Rowhammer Bug}
The increase of DRAM density has led to physically smaller cells, thus capable of storing smaller charges. As a result, cells have a lower noise margin, and cells can interact electrically with each other although they should be isolated. The so called \textit{Rowhammer bug} consists in the corruption of data, not in rows that are directly accessed, but rather in rows nearby the accessed one.

DRAM and CPU manufacturers have known the Rowhammer bug since at least 2012~\cite{bains2014row1,bains2014row2}. Hammering DRAM chips is a quality assurance tests applied to modules~\cite{al2005dram}. As refreshing DRAM cells consumes time, DRAM manufacturers optimize the refresh rate to the lowest rate that still works reliably.

The Rowhammer bug has recently been studied~\cite{Kim2014ISCA,Huang2012,Park2014} and the majority of off-the-shelf DRAM modules has been found vulnerable to bit flips using the \texttt{clflush} instruction.
The \texttt{clflush} instruction flushes data from the cache, forcing the CPU to serve the next memory access from DRAM.
Their proof-of-concept implementation frequently accesses and flushes two memory locations in a loop, causing bit flips in a third memory location.

Seaborn implemented Rowhammer exploits~\cite{SeabornBlackhat2015} in native code with the \texttt{clflush} instruction: a privilege escalation on a Linux system caused by a bit flip in a page table and an escape from the Google Native Client sandbox caused by a bit flip in indirect jumps.
As a countermeasure, the \texttt{clflush} instruction was removed from the set of allowed instructions in Google Chrome Native Client~\cite{SeabornBlackhat2015}.

\subsection{CPU Caches}\label{cpucaches}
A CPU cache is a small and fast memory inside the CPU hiding the latency of main memory by keeping copies of frequently used data. Modern Intel CPUs have three levels of cache, where L1 is the smallest and fastest cache and L3 the slowest and largest cache.
The L3 cache is an inclusive cache, \ie all data in L1 and L2 cache is also present in the L3 cache.
It is divided into one slice per CPU core, but shared, \ie cores can access all slices. The undocumented \textit{complex addressing} function that maps physical addresses to slices was recently reverse engineered~\cite{Maurice2015RAID,Inci2015iacr,Yarom2015iacr}. We used the results published by Maurice~\etal\cite{Maurice2015RAID}, shown in Table~\ref{fig:complex_addressing}. The table shows how address bits 6 to 32 are xor'd into one or two output bits $o_0$ and $o_1$. In case of a dual-core CPU, output bit $o_0$ determines to which of the two cache slices the physical address maps. In case of a quad-core CPU, output bits $o_1$ and $o_0$ determine the slice.

\begin{table}[!t]
\centering
\bgroup
\setlength{\tabcolsep}{0.1em}
\caption{Complex addressing function from~\cite{Maurice2015RAID}.}\label{fig:complex_addressing}
\scriptsize
\begin{tabular}{|p{0.9cm}|c||c|c|c|c|c|c|c|c|c|c|c|c|c|c|c|c|c|c|c|c|c|c|c|c|c|c|c|}
  \cline{3-29}
\multicolumn{2}{c|}{} & \multicolumn{27}{c|}{Address Bit}\\
\hline &  & 3 & 3 & 3 & 2 & 2 & 2 & 2 & 2 & 2 & 2 & 2 & 2 & 2 & 1 & 1 & 1 & 1 & 1 & 1 & 1 & 1 & 1 & 1 & 0 & 0 & 0 & 0 \\ 
 & & 2 & 1 & 0 & 9 & 8 & 7 & 6 & 5 & 4 & 3 & 2 & 1 & 0 & 9 & 8 & 7 & 6 & 5 & 4 & 3 & 2 & 1 & 0 & 9 & 8 & 7 & 6 \\ 
\hline\hline 2 cores & $o_0$ & $\xor$ &  & $\xor$ &  & $\xor$ & $\xor$ & $\xor$ & $\xor$ & $\xor$ &  & $\xor$ &  & $\xor$ &  & $\xor$ & $\xor$ & $\xor$ &  & $\xor$ &  & $\xor$ &  & $\xor$ &  &  &  & $\xor$ \\  
\hline\hline \multirow{2}{*}{4 cores}& $o_0$ & $\xor$ &  & $\xor$ &  & $\xor$ & $\xor$ & $\xor$ & $\xor$ & $\xor$ &  & $\xor$ &  & $\xor$ &  & $\xor$ & $\xor$ & $\xor$ &  & $\xor$ &  & $\xor$ &  & $\xor$ & \phantom{$\xor$} &  &  & $\xor$ \\ 
\cline{2-29} & $o_1$ &  & $\xor$ &  & $\xor$ & $\xor$ &  & $\xor$ &  & $\xor$ & $\xor$ & $\xor$ & $\xor$ & $\xor$ & $\xor$ &  & $\xor$ &  & $\xor$ &  & $\xor$ &  & $\xor$ &  &  & \phantom{$\xor$} &  $\xor$ &  \\ 
\hline
\end{tabular}
\egroup
\end{table}

Caches are organized in sets of multiple lines. The mapping from physical addresses to sets is fixed.
Addresses that map to the same set and slice are called \textit{congruent}.
To load a new line from memory, the \textit{replacement policy} decides which line to evict.
Intel has not disclosed the cache replacement policy of their CPUs. However, the replacement policies for some architectures have been reverse-engineered: Sandy Bridge has a pseudo-LRU replacement policy and Ivy Bridge a modification of the pseudo-LRU replacement policy~\cite{Wong2013}. Moreover, Ivy Bridge, Haswell and Skylake use adaptive cache replacement policies which only behave as pseudo-LRU in some situations~\cite{Qureshi2007}. These CPUs can switch the cache replacement policy frequently. 

\subsection{Cache Attacks and Cache Eviction}
Cache side-channel attacks exploit timing differences between cache hits and cache misses. 
Practical attacks on cryptographic algorithms have been explored thoroughly~\cite{2004-bernstein-cachetiming,2005-percival-cache}. There are two main types of cache attacks called Prime+Probe and Flush+Reload.
The Prime+Probe attack has been introduced by Percival~\cite{2005-percival-cache} and Osvik~\etal\cite{DBLP:conf/ctrsa/OsvikST06}. It determines activities of a victim process by repeatedly measuring the duration to access once every address in a set of congruent addresses, \ie a so-called eviction set.
Prime+Probe on the last-level cache enables cross-core cache attacks such as cross-VM attacks without shared memory~\cite{DBLP:conf/sp/IrazoquiES15,Liu2015}, covert channels~\cite{Maurice2015C5} and attacks from within sandboxed JavaScript~\cite{Oren2015CCS}.
Oren~\etal\cite{Oren2015CCS} and Liu~\etal\cite{Liu2015} compute the eviction set by adding addresses to the eviction set until eviction works.
Flush+Reload has been introduced by Gullasch~\etal\cite{DBLP:conf/sp/GullaschBK11} and Yarom and Falkner~\cite{DBLP:conf/uss/YaromF14}. It exploits shared memory between attacker and victim and is very fine-grained. Cache lines are flushed with the \texttt{clflush} instruction or using cache eviction~\cite{191010}.

Evicting data from the cache is just as crucial to cache attacks as it is for the Rowhammer attack.
Previous work either uses the \texttt{clflush} instruction or hand-crafted eviction loops.
Hund~\etal\cite{Hund2013} showed that data can be evicted by filling a large memory buffer the size of the cache. However, this is very slow and thus not applicable to fine-grained cache attacks or Rowhammer attacks. Using the reverse-engineered complex addressing function solves the problem of finding addresses that are congruent in the cache, but it leaves the non-trivial problem of finding access sequences to achieve high eviction rates while maintaining a low execution time.

\section{Cache Eviction Strategies}\label{sec:opt_cache_evict}

In this section, we describe how to find cache eviction strategies in a fully automated way for microarchitectures post Sandy Bridge. An \textit{eviction strategy} accesses addresses from an eviction set in a specific access pattern and can ideally be used as a replacement for \texttt{clflush}. \textit{Eviction set} is commonly defined as a set of congruent addresses. The access pattern defines in which order addresses from the eviction set are accessed, including multiple accesses per address.

An efficient eviction strategy can replace the \texttt{clflush} instruction in any cache attack and significantly improves cache attacks based on Prime+Probe, like JavaScript-based attacks~\cite{Oren2015CCS} or cross-VM cache attacks~\cite{Liu2015}. It also allows to replace the \texttt{clflush} instruction in a Rowhammer attack (see Section~\ref{sec:rowhammer_without_flush}).

The replacement policy of the CPU influences the size of the eviction set and the access pattern necessary to build an efficient eviction strategy.
For a pseudo-LRU replacement policy, accessing as many congruent locations as the number of ways of the L3 cache (for instance 12 or 16) once, evicts the targeted address with a high probability.
For adaptive cache replacement policies, an eviction strategy that is effective for one policy is likely to be ineffective for the other. Thus it is necessary to craft an eviction strategy that causes eviction for both policies and ideally does not introduce a significant timing overhead.

We distinguish between the following ways to generate an eviction strategy:
\begin{compactenum}
	\item \textit{Static eviction set and static access pattern}: uses information on cache slice function and physical addresses, and generates a pre-defined pattern in negligible time.
	Sections~\ref{sec:opt_cache_evict_static} and~\ref{sec:assumption_based_eviction} describe new efficient eviction strategies computed this way.
	\item \textit{Dynamic eviction set and static access pattern}: computes the eviction set in an automated way, without any knowledge of the system, \eg the number of cores.
A good access pattern that matches the replacement policy of the targeted system is necessary for a successful attack. Section~\ref{sec:assumption_based_eviction} describes this approach.
	\item \textit{Dynamic eviction set and dynamic access pattern}: automatically computes the eviction set and the access pattern based on randomness. This comes at the cost of performing a huge number of eviction tests, but it has the advantage to require almost no information on the system, and allows to implement fully automated online attacks for unknown systems. Section~\ref{sec:opt_cache_evict_dynamic} describes this approach.
	\item \textit{Static eviction set and dynamic access pattern}: uses a pre-defined eviction set, but a random pattern that is computed in an automated way. This is possible in theory, but it has no advantage over automatically testing static access patterns. We thus do not further investigate this approach. 
\end{compactenum}

We first describe a model to represent access patterns, given several parameters. To find a good eviction strategy for a given system, we define an offline and an online phase. In the offline phase, the attacker explores the parameter space to find the best eviction strategies for a set of controlled systems. The goal is to find a eviction strategy that matches the undocumented replacement policy the closest, including the possibility of policy switches. In the online phase, the attacker targets an unknown system, with no privileges.

\subsection{Cache Eviction Strategy Model}\label{sec:model_cache_strategies}
The success of a cache eviction strategy is measured by testing whether the targeted memory address is not cached anymore over many experiments, \ie average success rate. For such cases, we made the following three observations.

First, only cache hits and cache misses to addresses in the same cache set have a non-negligible influence on the cache, apart from cache maintenance and prefetching operations to the same cache set. We verified this by taking an eviction algorithm and randomly adding memory accesses that are not congruent. The eviction rate is the average success rate of the eviction function. It does not change by adding non-congruent accesses to an eviction strategy as long as the timing does not deviate. Thus, the eviction set only contains congruent addresses and the effectiveness of the eviction strategy depends on the \textit{eviction set size}.

Second, addresses are indistinguishable with respect to the cache. Thus, we represent access patterns as sequences of address labels $a_i$, \eg $a_1a_2a_3\ldots$. Each address label is set to a different address and thus for each time frame the sequence defines which address to access. A pattern $a_1a_2a_3$ is equivalent to any pattern $a_ka_la_m$ where $k\neq l\neq m$. If run in a loop, the number of \textit{different memory addresses} has an influence on the effectiveness on the eviction strategy.

Third, repeated accesses to the same address are necessary to keep it in the cache, as replacement policies can prefer to evict recently added cache lines over older ones. Changing the eviction sequence from $a_1a_2\ldots a_{17}$ to $a_1a_1a_2a_2\ldots a_{17}a_{17}$ reduces the execution time by more than $33\%$ on Haswell, and increases the eviction rate significantly if executed repeatedly, as the cache remains filled with our eviction set. However, we observed a diminishing marginal utility for the number of accesses to the same address. For all addresses we observed that after a certain number of accesses, further accesses do not increase and can even decrease the eviction rate. Thus, we describe eviction strategies as a loop over an eviction set of size $S$, where only a subset of $D$ addresses is \textit{accessed per round}. A parameter $L$ allows to make accesses \textit{overlap} for repeated accesses.

\begin{lstlisting}[float,
                   caption=Eviction loop for pattern testing.\vspace{-0.3cm},
                   label=lst:pattern_loop]
for (s = 0; s <= S-D; s += L)
  for (c = 0; c <= C; c += 1)
    for (d = 0; d <= D; d += 1)
      *a[s+d];
\end{lstlisting}

While testing all possible sequences even for very small sequence lengths is not possible in practical time (\cf Stirling numbers of second kind as a good estimate), a systematic exploration of influential parameters is possible. 
In theory, better eviction strategies may lie outside of this reduced search space. However using this method, we found eviction strategies that allowed us to successfully trigger bit flips using eviction-based Rowhammer (see Section~\ref{sec:rowhammer_without_flush}). 
To discuss and compare eviction strategies systematically, we use the following naming scheme in this paper to describe parametrized eviction strategies as depicted in Listing~\ref{lst:pattern_loop}. The eviction strategy name has the form $\mathcal{P}\mhyphen {C}\mhyphen {D}\mhyphen {L}\mhyphen {S}$, with
${C}$, the number of accesses to each memory address per loop round,
${D}$, the number of different memory addresses accessed per loop round,
${L}$, the step size/increment of the loop (for overlapping accesses),
and ${S}$, the eviction set size.
For instance, LRU-eviction is $\mathcal{P}\mhyphen 1\mhyphen 1\mhyphen 1\mhyphen {S}$ with an access sequence of $a_1a_2a_3\ldots a_S$.

\subsection{Offline Phase}\label{sec:opt_cache_evict_static}
In the offline phase, the attacker has at his disposal a set of machines and tries to learn the eviction strategy that matches the replacement policy the closest for each machine. While it is not strictly a reverse engineering of the replacement policy, by knowing the best eviction strategy, the attacker gains knowledge on the systems. In this phase, the attacker has no time constraints.

We discuss the evaluation in detail for the Haswell platform with a single DIMM in single channel mode. 
We explored the parameter space up to degree 6 in the dimensions of ${C}$, ${D}$ and ${L}$ and 23 different eviction set sizes each, in order to find eviction strategies that are fast and effective enough to perform Rowhammer attacks. Including the equivalent eviction strategies we evaluated a total of 18293 eviction strategies on 3 of our test platforms. We tested each eviction strategy in 20 double-sided Rowhammer tests with 2 million hammering rounds (\ie 80 million evictions per eviction strategy) and evaluated them using different evaluation criteria including eviction rate, runtime, number of cache hits and misses. The runtime was more than 6 days. The hammering was performed on a fixed set of physical addresses congruent to one specific cache set to allow for a fair comparison of the eviction strategies. Half of the evictions, \ie 40 millions, were used to measure eviction rate, cache hits and cache misses. The other half was used to measure the average execution time per eviction. We verified that the sample size is high enough to get reproducible measurements.

The number of bit flips is not suitable for the evaluation of a single eviction strategy, but only to determine whether and how cache hits, cache misses, the execution time and the eviction rate influence the probability of a bit flip.
Bit flips are reproducible in terms of the memory location, but the time and the number of memory accesses until a bit flip occurs again varies widely.
In order to measure the average number of bit flips for a eviction strategy, we would have to test every eviction strategy for several hours instead of minutes. This would increase the test time per machine to several weeks, and even then, it would not yield reproducible results, as it has been observed that the DRAM cells get permanently damaged if hammered for a long time~\cite{Kim2014ISCA}.

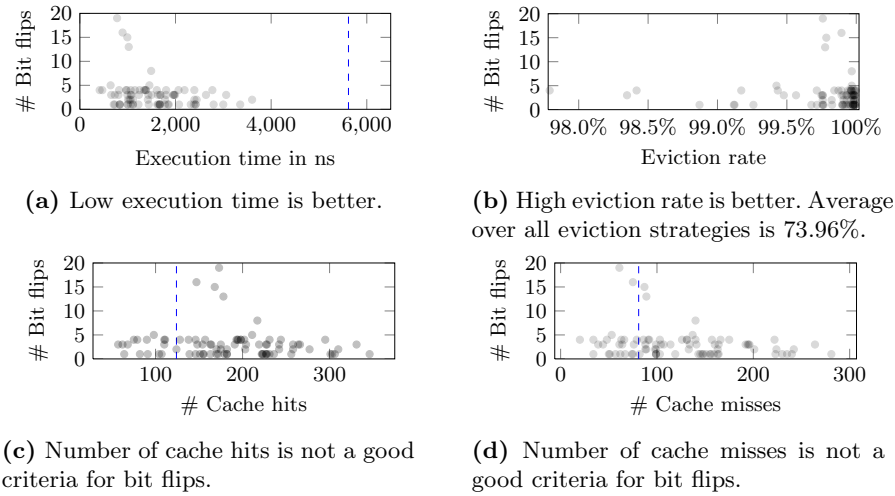
\begin{figure}[t!]
\centering
\begin{subfigure}{.46\hsize}
\centering
\begin{tikzpicture}[scale=0.9]
\pgfplotsset{every axis legend/.append style={at={(0.5,1.4)},anchor=north}}
\begin{axis}[
legend columns=4,
xlabel=Execution time in ns,
ylabel=\# Bit flips,
width=1.1\hsize,
ymin=0,
ymax=20,
xmin=0,
xmax=6500,
height=3cm,
]
\addplot+[color=black,mark options={solid,fill=black,opacity=0.15},only marks,mark size=1.5pt] table[x=time,y=sum] {time_flips.csv};
\addplot[dashed, blue] coordinates {(5617,0)(5617,20)};
\end{axis}
\end{tikzpicture}
\caption{Low execution time is better. \\ \qquad \qquad}
\label{fig:time_flips}
\end{subfigure}
\qquad
\begin{subfigure}{0.46\hsize}
\centering
\begin{tikzpicture}[scale=0.9]
\pgfplotsset{every axis legend/.append style={at={(0.5,1.4)},anchor=north}}
\begin{axis}[
legend columns=4,
xlabel=Eviction rate,
ylabel=\# Bit flips,
xticklabel style={ /pgf/number format/fixed, /pgf/number format/precision=5},
ymin=0,
ymax=20,
xmin=0.9778,
xmax=1.0002,
xtick={0.98,0.985,0.99,0.995,1},
xticklabels={$98.0\%$,$98.5\%$,$99.0\%$,$99.5\%$,$100\%$},
width=1.1\hsize,
height=3cm,
]
\addplot+[color=black,mark options={solid,fill=black,opacity=0.15},only marks,mark size=1.5pt] table[x=rate,y=sum] {eviction_rate_flips.csv};
\end{axis}
\end{tikzpicture}
\caption{High eviction rate is better. Average over all eviction strategies is $73.96\%$.}
\label{fig:eviction_rate_flips}
\end{subfigure}
\begin{subfigure}{0.45\hsize}
\centering
\begin{tikzpicture}[scale=0.9]
\pgfplotsset{every axis legend/.append style={at={(0.5,1.4)},anchor=north}}
\begin{axis}[
legend columns=4,
xlabel=\# Cache hits,
ylabel=\# Bit flips,
ymin=0,
ymax=20,
width=1.1\hsize,
height=3cm,
]
\addplot+[color=black,mark options={solid,fill=black,opacity=0.25},only marks,mark size=1.5pt] table[x=hits,y=sum] {hits_flips.csv};
\addplot[dashed, blue] coordinates {(124,0)(124,20)};
\end{axis}
\end{tikzpicture}
\caption{Number of cache hits is not a good criteria for bit flips.}
\label{fig:hits_flips}
\end{subfigure}
\qquad
\begin{subfigure}{0.45\hsize}
\centering
\begin{tikzpicture}[scale=0.9]
\pgfplotsset{every axis legend/.append style={at={(0.5,1.4)},anchor=north}}
\begin{axis}[
legend columns=4,
xlabel=\# Cache misses,
ylabel=\# Bit flips,
ymin=0,
ymax=20,
width=1.1\hsize,
height=3cm,
]
\addplot+[color=black,mark options={solid,fill=black,opacity=0.15},only marks,mark size=1.5pt] table[x=misses,y=sum] {misses_flips.csv};
\addplot[dashed, blue] coordinates {(81,0)(81,20)};
\end{axis}
\end{tikzpicture}
\caption{Number of cache misses is not a good criteria for bit flips.}
\label{fig:misses_flips}
\end{subfigure}
\caption{Relation between the number of bit flips and average execution time, cache hits and cache misses per eviction and the eviction rate of the corresponding eviction strategy measured in 40 million samples. One point per eviction strategy that caused a bit flip, others are omitted. The darker the more points overlay. Average over all eviction strategies shown as dashed line. Good eviction strategies have high eviction rates and low execution times.}
\label{fig:bitflips_comparison}
\end{figure}

High execution times are too slow to trigger bit flips and low execution times are useless without a good eviction rate.
The execution time of the eviction strategy is directly related to the number of memory accesses to the two victim addresses. Hence, it influences the probability of a bit flip directly. On our default configured Ivy Bridge notebook we observed bit flips even with execution times of 1.5 microseconds per hammering round, that is approximately 21,500 accesses per address within the specified total refresh interval of 64ms. This maps to the average periodic refresh interval \texttt{tREFI} by dividing 64ms by 8192~\cite{MicronTN-46-09}.
Double-sided rowhammering using \verb|clflush| takes only 60 nanoseconds on our Haswell test system, that is approximately 0.6 million accesses per address in 64ms. Figure~\ref{fig:time_flips} shows how bit flips are correlated with the eviction execution time.

The eviction rate has to be very high to trigger bit flips. Figure~\ref{fig:eviction_rate_flips} shows how many bit flips occurred at which eviction rate. We observe that $81\%$ of the bit flips occurred at an eviction rate of $99.75\%$ or higher and thus use this as a threshold for good eviction strategies on our Haswell system. Even though a bit flip may occur at lower eviction rates, the probability is significantly lower.

The eviction loop contributes to a high number of cache hits and cache misses, apart from the two addresses we want to hammer. We measure the number of cache hits and cache misses that occur during our test run using hardware performance counters through the Linux syscall interface \verb|perf_event_open|.
Cache hits have a negligible influence on the execution time and no effect on the DRAM. Cache misses increase the execution time and, if performed on a different row but in the same channel, rank and bank, additional DRAM accesses. However, Figures~\ref{fig:hits_flips} and~\ref{fig:misses_flips} show that both cache hits and cache misses do not impact the number of bit flips significantly, as the average for all eviction strategies is in the range of the eviction strategies that triggered a bit flip.

Thus, we thus use the eviction rate as a criteria for good eviction strategies, and among those eviction strategies, we prefer those with a lower average execution time. This method requires no access to any system interfaces and can be implemented in any language and execution environment that allows to measure time and perform arbitrary memory accesses, such as JavaScript.

\begin{table}[t]
\caption{The fastest 5 eviction strategies with an eviction rate above $99.75\%$ compared to \texttt{clflush} and LRU eviction on the Haswell test system.}\label{tab:eviction_systematic}
\csvautobookrtabular[respect percent=true]{eviction_rate_systematic.csv}
\end{table}

Table~\ref{tab:eviction_systematic} shows a comparison of the fastest 5 of these eviction strategies with an eviction rate above $99.75\%$ (see Figure~\ref{fig:eviction_rate_flips}) and \texttt{clflush} based rowhammering as well as the fastest LRU ($\mathcal{P}\mhyphen 1\mhyphen 1\mhyphen 1\mhyphen 20$) eviction strategy that achieves the same eviction rate. The best two eviction strategies are $\mathcal{P}\mhyphen 5\mhyphen 2\mhyphen 2\mhyphen 18$ and $\mathcal{P}\mhyphen 2\mhyphen 2\mhyphen 1\mhyphen 17$, both with an execution time around 180 nanoseconds.

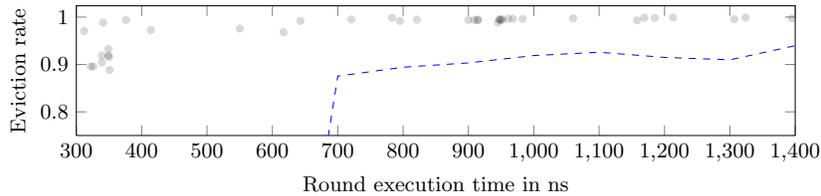
\begin{figure}[t]
\centering
\begin{tikzpicture}[scale=0.9]
\pgfplotsset{every axis legend/.append style={at={(0.5,1.4)},anchor=north}}
\begin{axis}[
legend columns=4,
xlabel=Round execution time in ns,
ylabel=Eviction rate,
width=\hsize,
ymin=0.75,
xmin=300,
xmax=1400,
height=3.5cm,
]
\addplot+[color=black,mark options={solid,fill=black,opacity=0.15},only marks,mark size=1.5pt] table[x=time,y=rate] {time_flips_ivy.csv};
\addplot+[color=blue,dashed,no marks] table[x=time,y=rate] {time_flips_ivy_avg.csv};
\end{axis}
\end{tikzpicture}
\caption{Average execution time and eviction rate per eviction strategy on Ivy Bridge measured in 40 million samples per eviction strategy. One point per eviction strategy that caused a bit flip, others are omitted. The darker the more points overlay. Average over all eviction strategies shown as dashed line.}
\label{fig:time_flips_ivy}
\end{figure}

Accessing each address in the eviction set only once (LRU eviction) is far from optimal for cache attacks and impractical for Rowhammer.
Although counterintuitive, adding more accesses to the eviction loop will lower the overall execution time.
We can observe this for instance by comparing the eviction strategies $\mathcal{P}\mhyphen 1\mhyphen 1\mhyphen 1\mhyphen 20$ and $\mathcal{P}\mhyphen 4\mhyphen 2\mhyphen 2\mhyphen 20$. While both access the same set of 20 addresses, the latter one performs 4 times as many memory accesses, yet its execution time is only one third. Comparing the best eviction strategy we found to LRU eviction as described in previous work, performs only as good if the set size is at least $S=25$, increasing the average execution time 9 times higher than the one of the best eviction strategy we found. On the other hand, the eviction set size in previous work is typically specified as $S=17$. For $\mathcal{P}\mhyphen 1\mhyphen 1\mhyphen 1\mhyphen 17$ we measured an eviction rate of $74.5\%$ and even then a $1.7$ times higher execution time than with the best eviction strategy we found. 
This shows that the eviction strategies we found are a significant improvement over previously published eviction methods.

We performed the same evaluation for the other architectures. The distribution of bit flips on our Ivy Bridge test system relative to eviction rate and execution time is shown in Figure~\ref{fig:time_flips_ivy}. Most bit flips occurred at eviction rates above $99\%$. The fastest 5 of these eviction strategies are shown in Table~\ref{tab:eviction_systematic_ivy_skylake} in comparison with \texttt{clflush} and the fastest LRU ($\mathcal{P}\mhyphen 1\mhyphen 1\mhyphen 1\mhyphen 15$) eviction strategy.

\begin{table}[t]
\caption{\texttt{clflush} and LRU eviction compared to the fastest 5 eviction strategies above $99\%$ eviction rate on the Ivy Bridge test system (left) and compared to the fastest 5 eviction strategies above $99.9\%$ eviction rate on the Skylake DDR4 test system (right).}\label{tab:eviction_systematic_ivy_skylake}
\csvautobookrtabular[respect percent=true]{eviction_rate_systematic_ivy.csv} \hfill
\csvautobookrtabular[respect percent=true]{eviction_rate_systematic_skylake.csv}
\end{table}

According to our measurements the complex addressing function on Skylake is not the same as in Haswell, but it can be trivially derived from the reverse engineered 8-core function. We again found that LRU eviction performs much worse than the best eviction strategy we found as shown in Table~\ref{tab:eviction_systematic_ivy_skylake}.
\subsection{Online Phase}
In the online phase, the attacker targets an unknown system. In particular, microarchitecture and number of CPU cores are unknown to the attacker. The attacker has the knowledge gained from the offline phase at his disposal. However, he has no privilege on the victim's machine and no time to run the extensive search from the offline phase. The online phase consists in two attacks: an assumption-based attack, and a fall-back attack in case the first one does not work. In both cases the attack is based on a series of timing attacks and no access to specific system interfaces is necessary.
\subsubsection{Assumption-based Attack}\label{sec:assumption_based_eviction}
The attacker first tests whether the targeted system resembles a system tested in the offline phase, by performing timing attacks. No access to syscalls or system interfaces is required for this step. The attacker defines a threshold eviction rate based on the results from the offline phase (for instance $99.75\%$) and searches for eviction strategies above this threshold on the system under attack. By testing a set of eviction strategies from the offline phase, the attacker learns whether the architecture of the system under attack resembles an architecture from the offline phase. In this case the best eviction strategy for the system under attack is within the set of eviction strategies previously tested. The number of eviction strategies to test is as low as the number of targeted CPU architectures and thus it only takes a few seconds to compute.

\begin{figure}[t]
\centering
{\scriptsize
\texttt{
\vspace{-0.5em}
\\
0123 0123 0123 0123 1032 1032 1032 1032 2301 2301 2301 2301 3210 3210 3210 3210\\
1032 1032 1032 1032 0123 0123 0123 0123 3210 3210 3210 3210 2301 2301 2301 2301\\
2301 2301 2301 2301 3210 3210 3210 3210 0123 0123 0123 0123 1032 1032 1032 1032\\
3210 3210 3210 3210 2301 2301 2301 2301 1032 1032 1032 1032 0123 0123 0123 0123\\
}
}
\caption{Slice patterns for 64-byte offsets on 4KB pages on a 4-core system. An attacker can derive which addresses map to the same cache slice. Substituting 2 by 0 and 3 by 1 gives the slice pattern for 2-core systems.}
\label{fig:cache_patterns}
\end{figure}

The eviction set can be computed in a static or dynamic way.
Without any further assumptions we can run modified versions of the algorithms by Oren~\etal\cite{Oren2015CCS} or Liu~\etal\cite{Liu2015}. Instead of the $\mathcal{P}\mhyphen 1\mhyphen 1\mhyphen 1$ access pattern they implement, we use one of the suspected eviction strategies to build a dynamic assumption-based algorithm. This improves the success rate of their algorithms on recent architectures.
However, we make additional assumptions to reduce the execution time to a minimum and build a static assumption-based algorithm. One assumption is that large arrays are allocated on large pages, as has been observed before~\cite{Gruss2015esorics}. Based on this assumption we can use the complex addressing function from Table~\ref{fig:complex_addressing} to determine the slice patterns for 4KB and 2MB pages as shown in Figure~\ref{fig:cache_patterns}. These distinct patterns in the mapping from physical addresses to cache slices depend only on the number of cache slices and are the same for Intel CPUs since the Sandy Bridge architecture. The algorithm by Oren~\etal\cite{Oren2015CCS} or Liu~\etal\cite{Liu2015} finds only addresses in the same cache slice and cache set. We use it to build an eviction set of 2MB-aligned congruent addresses in the same slice. Subsequent eviction set computations are performed statically based on the complex addressing function and the identified 2MB offsets.

\subsubsection{Fall-back Attack}\label{sec:opt_cache_evict_dynamic}
If the assumption-based phase does not work on a system under attack, \eg because the unknown system is none of the systems tested in the offline phase, the attacker runs a fall-back phase to find an eviction strategy that is sufficient to trigger a bit flip with Rowhammer.

Oren~\etal\cite{Oren2015CCS} and Liu~\etal\cite{Liu2015} compute a dynamic eviction set with a static access pattern $\mathcal{P}\mhyphen 1\mhyphen 1\mhyphen 1$. We extend their algorithms to compute eviction strategies with dynamic eviction sets and dynamic access patterns.
In the first step, we continuously add addresses to the eviction strategy multiple times to create eviction strategies with multiple accesses to the same address. We know that the eviction strategy is large enough as soon as we can clearly measure the eviction of the target physical address. 
In a second step, when the eviction rate is above the attacker chosen threshold, eviction addresses that do not lower the eviction rate are removed by replacing them with other addresses that are still in the eviction set. Thus, the number of memory accesses does not decrease, but the eviction set is minimized. This decreases the number of cache misses and thus the execution time. Finally, we randomly remove accesses that do not decrease the eviction rate and do not increase the execution time. This again decreases the number of unnecessary cache hits and thus the execution time.

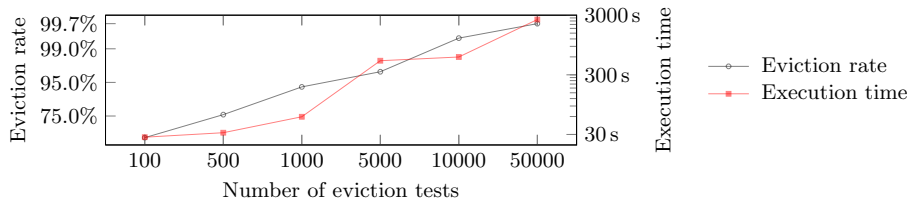
\begin{figure}[t!]
\vspace{-0.3cm}
\centering
\begin{tikzpicture}[scale=0.9]
\pgfplotsset{every axis legend/.append style={at={(0.72\hsize,0.5)},anchor=west}}
\begin{semilogyaxis}[
legend style={draw=none},
legend cell align=left,
xlabel=Number of eviction tests,
ylabel=Eviction rate,
ymin=0.2,
ymax=100,
xmin=0.5,
xmax=6.5,
ytick={25,5,1,0.3},
yticklabels={$75.0\%$,$95.0\%$,$99.0\%$,$99.7\%$},
y dir=reverse,
xticklabels={,,100,500,1000,5000,10000,50000},
ytick pos=left,
scaled y ticks = false,
width=.7\hsize,
height=3.5cm,
]
\addplot+[color=black,opacity=0.5,mark options={solid,draw=black,fill=black,opacity=0.5},mark=o,mark size=1pt] table[x=cached,y=rate] {dynamic_eviction_rate.csv};
\addlegendentry{~Eviction rate}
\addplot+[color=red,opacity=0.5,mark options={solid,draw=red,fill=red,opacity=0.5},mark=square*,mark size=1pt,draw=none] coordinates {(200,200)};
\addlegendentry{~Execution time}
\end{semilogyaxis}
\begin{semilogyaxis}[
legend style={draw=none},
legend cell align=left,
xlabel=,
ylabel=Execution time,
ymin=20,
ymax=3000,
xmin=0.5,
xmax=6.5,
scaled y ticks = false,
ytick={30,300,3000},
yticklabels={30\,s,300\,s,3000\,s},
ytick pos=right,
yticklabel pos=right,
trim axis right,
xticklabels={,,,,,},
width=.7\hsize,
height=3.5cm,
]
\addplot+[color=red,opacity=0.5,mark options={solid,draw=red,fill=red,fill=red,opacity=0.5},mark=square*,mark size=1pt] table[x=cached,y=time] {dynamic_eviction_rate.csv};
\end{semilogyaxis}
\end{tikzpicture}
\caption{The eviction rate and execution time of the dynamic eviction strategy when implementing the \texttt{cached(p)} function with $n$ eviction tests.}
\label{fig:cached_vs_eviction}
\end{figure}

The resulting eviction strategy can neither access less addresses nor can any duplicate accesses be removed without lowering the eviction rate. They thus perform similarly to statically computed eviction strategies.
The result of the algorithm is a series of accesses that fulfill the eviction rate threshold chosen by the attacker and that has a low execution time on the system under attack.
If the threshold was set high enough so that bit flips are likely to occur in practice, the eviction strategy found by the fall-back algorithm can be used for an attack.

The algorithm uses a function \verb|cached(p)| that tries to evict a target address \verb|p| using the current eviction strategy and set and decides whether \verb|p| is cached or not based on the access time. The quality of the solution depends on the number of tests that are performed in this function. The function only returns true, if an eviction rate below the attacker defined threshold is measured. A higher number of tests increases the execution time and the accuracy of this binary decision. Figure~\ref{fig:cached_vs_eviction} shows how the number of tests influences the eviction rate and the execution time of the resulting eviction strategy. If a high eviction rate is necessary, the execution time of the algorithm is can exceed 40 minutes. Thus, our algorithm can precompute a working eviction strategy once and subsequent eviction set computations are done with the fixed eviction strategy within seconds.

\section{Implementation of eviction-based Rowhammer}\label{sec:rowhammer_without_flush}
We now perform Rowhammer attacks using the eviction strategies from Section~\ref{sec:opt_cache_evict} instead of \texttt{clflush} in different scenarios. First, we demonstrate that it is possible to trigger bit flips in the same conditions as in the existing attacks where an attacker is able to execute native code on the system under attack. We then show that given knowledge about the physical addresses, it is possible to trigger
bit flips even from a remote website using JavaScript. In a third step, we show that
the full Rowhammer attack is possible from a remote website using JavaScript
without any additional information on the system.

\subsection{Rowhammer in Native Code}\label{sec:rowhammer_native}
We extended the \verb|double_sided_rowhammer| program by Dullien~\cite{SeabornBlackhat2015} by using the best eviction strategy we have found.
The two \texttt{clflush} instructions were first replaced by the eviction code described in Section~\ref{sec:model_cache_strategies}, with parameters for a $\mathcal{P}\mhyphen 2\mhyphen 2\mhyphen 1$ eviction strategy. The eviction sets are either precomputed statically using the physical address mapping and the complex addressing function in Table~\ref{fig:complex_addressing}, or using a dynamic eviction strategy computation algorithm.

This way, we were able to reproducibly flip bits on our Sandy Bridge and Ivy Bridge test machine using different eviction strategies when running with the Samsung DDR3 RAM and our Skylake test machine when running with the Crucial DDR4 RAM.
The machines were operated in default configuration.

On our Haswell test machine we were not able to reproducibly flip bits with the default settings, not even with the \texttt{clflush} instruction.
However, the BIOS configuration allows setting a custom refresh rate by setting the average periodic refresh interval \texttt{tREFI}.
We had to increase the \texttt{tREFI} value from 6,549 to over 19,000 just to be able to trigger bit flips \textit{with} the \verb|clflush| instruction.
The refresh interval is a typical parameter used by computer gaming enthusiasts and the overclocking community to increase system performance. However, while this might also be an interesting target group, we rather want to analyze the influence of the refresh interval on the applicability of the Rowhammer attack using cache eviction and the Rowhammer attack in JavaScript. Kim~\etal\cite{Kim2014ISCA} observed that the refresh interval directly influences the number of bit flips that occur and that below a module dependent \texttt{tREFI} value no bit flips occur. We will show that their observation also applies to Rowhammer with cache eviction and Rowhammer in JavaScript.

Lowering the refresh interval is not part of an actual attack. Existing work has already examined the prevalence of the Rowhammer and found that $85\%$ of the DDR3 modules examined are susceptible to Rowhammer bit flips~\cite{Kim2014ISCA}. Also in our case only the modules of the Haswell test system and the G.Skill DIMMs in the Skylake test system were not susceptible to Rowhammer bit flips at default settings, whereas it was possible to induce Rowhammer bit flips in the other three DIMMs at default settings. Thus, our results do not contradict previous estimates and we must assume that millions of systems are still vulnerable.

Rowhammer with eviction in native code revives the Google Native Client exploit~\cite{SeabornBlackhat2015} that allows privilege escalation in Google Chrome. The \texttt{clflush} instruction has been blacklisted to solve this vulnerability, however, this is ineffective and a sandbox escape is still possible, as we can trigger bit flips in Google Native Client based on eviction.

\subsection{Rowhammer in JavaScript}\label{sec:rowhammer_js}
Triggering the Rowhammer bug from JavaScript is more difficult as JavaScript has no concept of virtual addresses or pointers and no access to physical address mappings.
We observed that large typed arrays in JavaScript in all recent Firefox and Google Chrome versions on Linux are allocated 1MB aligned and use anonymous 2MB pages when possible. The reason for this lies in the memory allocation mechanism implemented by the operating system. Any memory allocation in a comparable scripting language and environment will also result in the allocation of anonymous 2MB pages for large arrays.

\begin{figure}[t]
\centering
\begin{tikzpicture}[scale=0.9]
\pgfplotsset{every axis legend/.append style={at={(0.5,1.4)},anchor=north}}
\begin{axis}[
legend columns=4,
xlabel=Page index,
ylabel={Latency in $\mu s$},
ymin=0,
xmin=0,
xmax=3072,
xtick={0,512,1024,1536,2048,2560,3072},
width=\hsize,
height=3.5cm,
]
\addplot+[color=black,mark options={solid,draw=black,fill=black},mark size=0.6pt] table[x=page,y=time] {page_time.csv};
\end{axis}
\end{tikzpicture}
\caption{Access latency of 4KB aligned addresses in a large array in JavaScript. Pagefaults cause the latency peaks at the start of the 2MB pages.}
\label{fig:page_timing}
\end{figure}
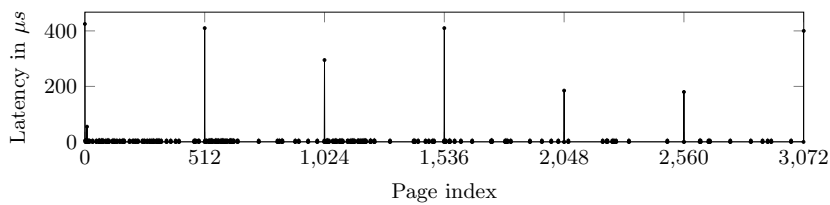

By performing a timing attack similar to the one performed by Gruss~\etal\cite{Gruss2015esorics}, we can determine the 2MB page frames in the browser. In this attack we iterate over an array and measure the access latency. The latency peaks during memory initialization are caused by the pagefaults that occur with the start of each new 2MB page, as shown in Figure~\ref{fig:page_timing}. This also works in recent browser versions with a reduced timer resolution as suggested by Oren~\etal\cite{Oren2015CCS} and added to the HTML5 standard by the W3C~\cite{2015W3C}. Thus, we know the lowest 21 bits of the virtual and physical address by knowing the offset in the array.

As a first proof-of-concept we reproduced bit flips in JavaScript in Firefox by hammering the exact physical addresses as in native code. In order to do this we built a tool to translate physical to virtual addresses for another process. To compute the eviction sets we use the assumption-based algorithm from Section~\ref{sec:opt_cache_evict_static}. We observed that simple memory accesses as in our native code implementation are not optimized out by the just-in-time-compiler.

The final JavaScript-based attack does not require any outside computation and thus, runs entirely without user interaction in the browser. It exploits the fact that large typed arrays are allocated on 2MB pages. Thus, we know that each 2MB region of our array is divided into 16 row offsets of size 128KB (depends on the lowest row index bit). We can now perform double-sided hammering in these 2MB regions to trigger a bit flip within the 2MB region or amplified single-sided hammering on the outer two rows of every 2MB pages to induce a bit flip in another physical 2MB region.
The result is the first hardware-fault attack implemented in JavaScript on a remote website.

\subsection{Attack Evaluation}
\begin{figure}[t!]
\centering
\begin{tikzpicture}[scale=0.9]
\pgfplotsset{every axis legend/.append style={at={(0.85\hsize,0.8)},anchor=north}}
\begin{semilogyaxis}[
legend style={draw=none},
legend cell align=left,
xlabel=Refresh interval in ns,
ylabel=Bit flips,
ymin=100,
width=0.8\hsize,
height=3.8cm,
]
\addplot+[unbounded coords=jump,color=black,mark options={solid,draw=black,fill=black},mark size=0.6pt] table[x=Refresh,y=Flush] {haswell_flips.csv};
\addlegendentry{~Flush (native)}
\addplot+[unbounded coords=jump,color=green,mark options={solid,draw=black,fill=black},mark size=0.6pt] table[x=Refresh,y=Evict] {haswell_flips.csv};
\addlegendentry{~Evict (native)}
\addplot+[unbounded coords=jump,color=red,mark options={solid,draw=black,fill=black},mark size=0.6pt] table[x=Refresh,y=JavaScript] {haswell_flips.csv};
\addlegendentry{~Evict (JavaScript)}
\end{semilogyaxis}
\end{tikzpicture}
\caption{Number of bit flips within 15 minutes on a fixed address pair for different values for the average periodic refresh interval \texttt{tREFI} on Haswell in three different setups.}
\label{fig:prob_bitflips_haswell}
\end{figure}
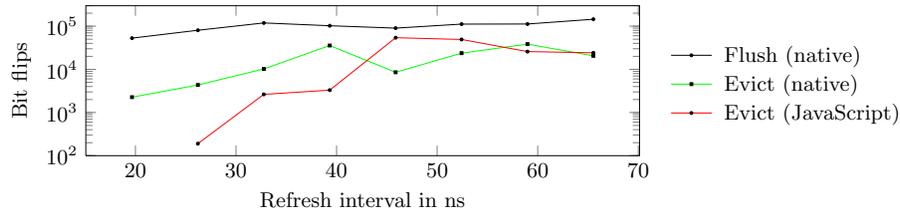

As described by Kim~\etal\cite{Kim2014ISCA} not all addresses in a DRAM are equally susceptible to bit flips. Therefore, to provide a fair comparison of the different techniques, we measured the number of bit flips for a fixed address pair already known to be susceptible.
Figure~\ref{fig:prob_bitflips_haswell} shows how different refresh rates influence the number of bit flips for a fixed time interval in different setups. 
The system was under slight usage during the tests (browsing, typing in an editor, etc.).
We see that the \texttt{clflush} instruction yields the highest number of bit flips.
If the refresh interval was set to a value where bit flips can be triggered using \texttt{clflush}, they can be triggered using native code eviction as well.
To trigger bit flips in JavaScript, a slightly higher refresh interval was necessary. Again, it depends on the particular DIMM whether the refresh interval is chosen correctly so that no bit flips occur.

The probability for bit flips in JavaScript is slightly lower than in native code, as native code is slightly faster. However, if a machine is vulnerable to our native code implementation it is likely vulnerable using our JavaScript implementation as well. While these plots were obtained on the Haswell machine, we were also able to trigger bit flips on our Ivy Bridge laptop with default settings from JavaScript. However, as the Laptop BIOSes did not allow to set the refresh interval \texttt{tREFI} directly, we could not obtain a comparable plot.

While DDR4 was assumed to have countermeasures against rowhammering, countermeasures are not part of the final DDR4 standard~\cite{Aichinger2015}. Using the Crucial DDR4 DIMMs we even were able to induce bit flips at default system settings and with the most recent BIOS version, after applying the functions reverse engineered by Pessl~\etal\cite{Pessl2015}. On the G.Skill DDR4 DIMMs we could only induce bit flips at an increased refresh interval. Thus, even on these very recent and up-to-date systems Rowhammer countermeasures have not been implemented in hardware and those implemented in software are ineffective. Whether a system is vulnerable to Rowhammer-based attacks still crucially depends on the refresh interval chosen by DIMM.

\section{Discussion and Related Work}\label{sec:discussion}

\subsection{Building an Exploit With Rowhammer.js}

Existing exploits assume that a page table is mapped in a row between two rows occupied by the attacker. However, we observed that this situation rarely occurs in practice. The operating system prefers to use large pages to reduce the pressure on the TLB. To make the organization and changes to physical address mappings easier the operating system will also group small pages into the same organizational physical frames. Page tables are only allocated between two user pages in a near-out-of-memory situation. Thus, the exploits allocate almost all system memory to enforce such a situation~\cite{SeabornBlackhat2015}. However, swapping is enabled by default in all major operating systems and thus the system will be severely unresponsive due to swapping. In our proof-of-concept exploit, we perform ``amplified single-sided hammering''. By hammering two adjacent rows we increase the probability for a bit flip in a surrounding row significantly compared to single-sided hammering. This allows to induce bit flips even across the borders of physically coherent 2MB regions with a high probability. As we already have been able to trigger bit flips in JavaScript we will only focus on how to manipulate a page table similar to previous exploits~\cite{SeabornBlackhat2015}. The attacker can repeat any step of the attack as long as necessary to be successful.

In the first step, the exploit locates an exploitable bit flip as described in Section~\ref{sec:rowhammer_js}, \ie a bit flip in the $\frac{1}{3}$ of the page table bits that are used for physical addresses. An exploitable bit flip changes an address bit in a page table that is in an adjacent 2MB region. We have found such bit flips on our all our test machines. In the second step, the exploit script releases all pages but the two that have previously been hammered and the ones that are required for cache eviction. Thus, also the page that contained the bit flip is released.
Allocating arrays requires the browser to reserve virtual memory regions and to map them to physical memory upon the first access. The attacker determines the largest array size that still triggers the allocation of a page table in a timing attack (see~\ref{sec:rowhammer_js}). The array size was 1MB on all our test systems. We only access and thus allocate one 4KB page per 1MB array and thus 2 user pages per page table. The probability to place a group of page tables in the targeted 2MB region is $\approx\frac{1}{3}$. In the third step, the exploit script triggers the bit flip again and may find that its own memory mappings changed. With a chance of $\approx\frac{1}{3}$ the memory mapped is now one of the attackers page tables. The attacker can now change mapped addresses in that page table and if successful, has gained full access to the physical memory of the system. Our proof-of-concept works on recent Linux systems with all recent versions of Firefox and it does not require a near-out-of-memory situation. It does not work in Google Chrome due to the immediate allocation of all physical memory for an allocated 1MB array after a single access.

\subsection{Limitations}
In JavaScript we use 2MB pages to find congruent addresses and adjacent rows efficiently. If the operating system does not provide 2MB pages, we cannot perform double-sided or amplified single-sided hammering. However, the probability of a bit flip with single-sided hammering is significantly lower. 
Exploiting double-sided hammering with 2MB pages is not possible because we can then only induce bit flips in our own memory.
Thus, an attack is only possible with amplified single-sided hammering to induce a bit flip in an adjacent row in an adjacent 2MB page. There is only a limited number of such rows in a system.
Still the search for an exploitable bit flip can easily take several hours, especially as the probability of a bit flip in JavaScript is lower than in native code. Furthermore, if we cannot guess the best eviction strategy for the system, it will take up to an hour of precomputations to find a good eviction strategy. The victim has to stay on the website for the duration of the attack.
While this was the case in our proof-of-concept attack it is less realistic for a real-world attack.

\subsection{Countermeasures}\label{sec:countermeasures}
The operating system allocates memory in large physical memory frames (often 2MB) for reasons of optimization. Page tables, kernel pages and user pages are not allocated in the same memory frame, unless the system is close to out-of-memory (\ie allocating the last few kilobytes of physical memory). Thus, the most efficient Rowhammer attack (double-sided hammering) would not possible if the operating system memory allocator was less aggressive in near-out-of-memory situations. Preventing (amplified) single-sided hammering is more difficult, as hammering across the boundaries of a 2MB region is possible.

To fully close the attack vector for double-sided hammering, we also have to deal with read-only shared code and data, \ie shared libraries. If the attacker hammers on a shared library, a fault can be induced in this library. Therefore, shared libraries should not be shared over processes that run at different privilege levels or under different users. As a consequence, the attacker would be unable to escape from a sandbox or gain access to a higher privilege level using \verb|clflush| or eviction-based Rowhammer.

Kim~\etal\cite{Kim2014ISCA} proposed several countermeasures which should be implemented for new DRAM modules, including increasing the refresh rate. However, this would cause significant performance impacts. BIOS updates supplied so far only double the refresh rate, which is insufficient to prevent attacks on all DRAM modules. Moreover, many users to not update the BIOS unless it is unavoidable.

Pseudo Target Row Refresh (pTRR) and Target Row Refresh (TRR) are features that refresh neighboring rows when the number of accesses to one row exceeds a threshold. They have less overhead compared to double the refresh rate. Although TRR has been announced as implemented in all DDR4 modules it has been removed from the final DDR4 standard. Manufacturers can still choose to implement it in their devices, but if the memory controller does not support it, it has no effect. 

Error-correcting code (ECC) memory is often mentioned as a countermeasure against Rowhammer attacks. However, recent work shows that it cannot reliably protect against Rowhammer attacks.cases~\cite{Aichinger2015b,Lanteigne2016}.

At the software level, one proposed countermeasure is the detection using hardware performance counters~\cite{Herath2015,Payer2016,FlushFlush,Aweke2016}. The excessive number of cache references and cache hits allows to detect on-going attacks. However, this countermeasure can suffer from false positives, so it needs further evaluation before it can be brought to practice.

\subsection{Related Work}
The initial work by Kim~\etal\cite{Kim2014ISCA} and Seaborn's~\cite{SeabornBlackhat2015} root exploit made the scientific community aware of the security implications of a Rowhammer attack. However, to date, there have been very few other publications, focusing on different aspects than our work. 
Barbara Aichinger~\cite{Aichinger2015} analyzed Rowhammer faults in server systems where the problem exists in spite of ECC memory. She remarks that it will be difficult to fix the problem in the millions or even billions of DDR3 DRAMs in server systems.
Rahmati~\etal\cite{Rahmati2015} have shown that bit flips can be used to identify a system based on the unique and repeatable error pattern that occurs at a significantly increased refresh interval.
Our paper is the first to examine how to perform Rowhammer attacks based on cache eviction.\footnote{A draft of this paper was published online since July 24, 2015.}
Our cache eviction techniques facilitated cache side-channel attacks on ARM CPUs~\cite{LippGSM15}.
Concurrent and independent work by Aweke~\etal\cite{Aweke2016} has also demonstrated bit flips without \texttt{clflush} on a Sandy Bridge laptop. They focus on countermeasures, whereas we focus on attacking a wider range of architectures and environments.

\section{Future Work}\label{sec:future}
While we only investigated the possibility of a JavaScript Rowhammer attack in Firefox and Google Chrome on Linux, the attack exploits fundamental concepts that are inbuilt in the way hardware and operating system work. Whenever the operating system uses 4KB pages, page tables are required and at latest allocated when one of the 4KB pages belonging to this page table is accessed. Thus, the operating system cannot prevent that $\frac{1}{3}$ of memory is allocated for page tables. The same attack approach could be applied to hypervisors that allocate 4KB pages to virtual machines, even if they applies similar allocation mechanisms as the Linux kernel. While it might seem unreasonable and not realistic that hypervisors allocate 4KB pages, it in fact makes cross-VM page deduplication easier. According to Barresi~\etal\cite{Barresi2015}, page deduplication is in fact still widely used in public clouds. Our work opens the possibility for further investigation on whether page deduplication in fact is not only a problem for security and privacy of virtual machines, but a security problem for the hypervisor itself.

\section{Conclusion}\label{sec:conclusions}
In this paper, we presented Rowhammer.js, an implementation of the Rowhammer attack using fast cache eviction to trigger the Rowhammer bug with only regular memory accesses.
It is the first work to investigate eviction strategies to defeat complex cache replacement policies. This does not only enable to trigger Rowhammer in JavaScript, it also benefits research on cache attacks as it allows to perform attacks on recent and unknown CPUs fast and reliably. Our fully automated attack runs in JavaScript through a remote website and can gain unrestricted access to systems. The attack technique is independent of CPU microarchitecture, programming language and execution environment.

The majority of DDR3 modules are vulnerable and DDR4 modules can be vulnerable too. Thus, it is important to discover all Rowhammer attack vectors. Automated attacks through websites pose an enormous threat as they can be performed on millions of victim machines simultaneously.

\section{Acknowledgments}
We would like to thank our shepherd Stelios Sidiroglou-Douskos and our anonymous reviewers for their valuable comments and suggestions.
We would also like to thank Mark Seaborn, Thomas Dullien, Yossi Oren, Yuval Yarom, Barbara Aichinger, Peter Pessl and Raphael Spreitzer for feedback and advice.

\noindent\begin{tabular}{m{\dimexpr 0.13\hsize} m{2pt} m{\dimexpr 0.87\hsize-5\tabcolsep-2pt}}
\vspace*{1mm}\includegraphics[width=\hsize]{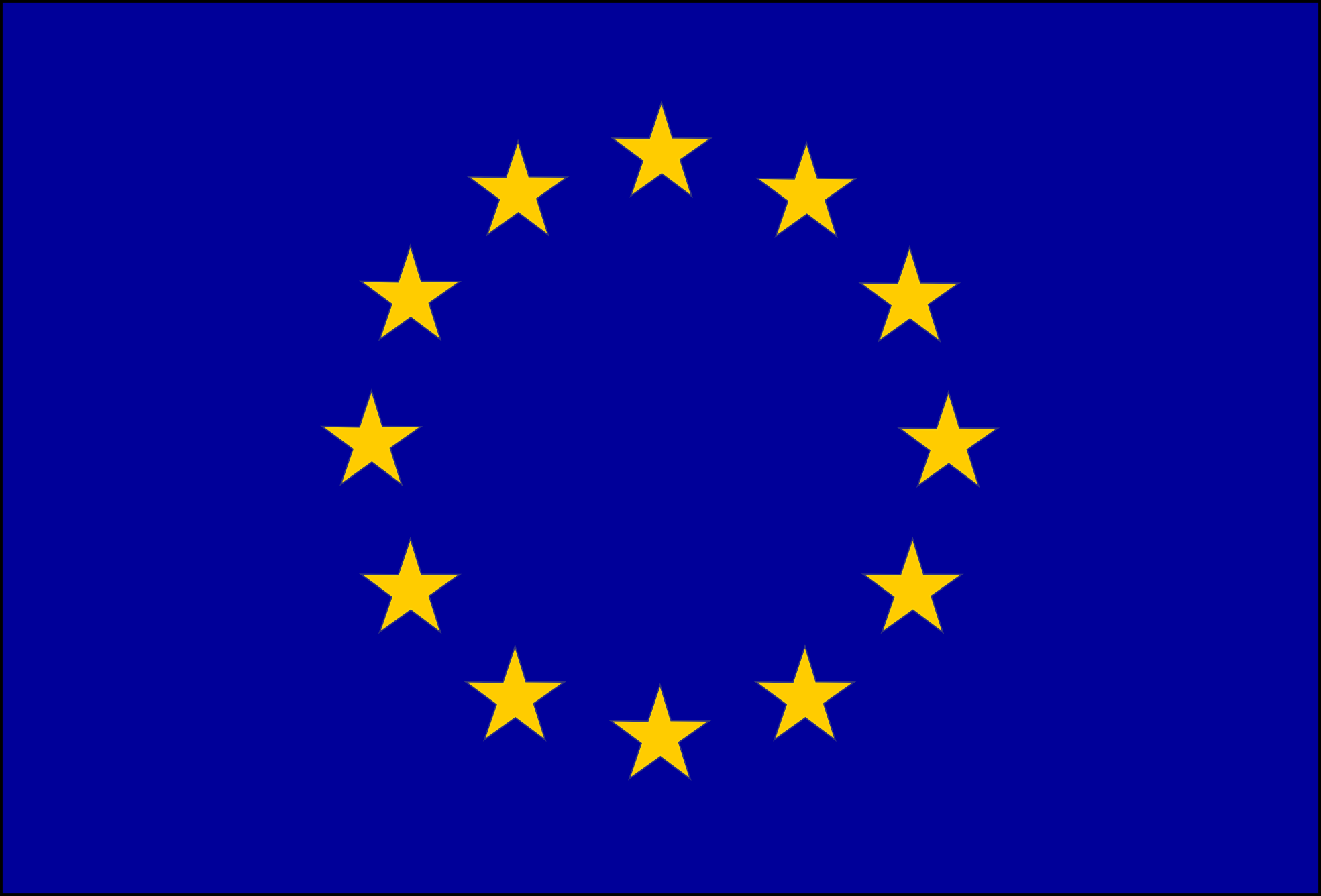}& &Supported by the EU Horizon 2020 programme under GA No. 644052 (HECTOR), the EU FP7 programme under GA No. 610436 (MATTHEW), the Austrian Research Promotion Agency (FFG) \quad \quad \quad \vspace{-\baselineskip}
\end{tabular} \\
and Styrian Business Promotion Agency (SFG) under GA No. 836628 (SeCoS), and Cryptacus COST Action IC1403.

{\footnotesize \bibliographystyle{splncs}
\bibliography{bibliography-short}}

\begin{thebibliography}{10}
\providecommand{\url}[1]{\texttt{#1}}
\providecommand{\urlprefix}{URL }

\bibitem{Aichinger2015}
Aichinger, B.: {DDR memory errors caused by Row Hammer}. In: HPEC'15 (2015)

\bibitem{Aichinger2015b}
Aichinger, B.: {Row Hammer Failures in DDR Memory}. In: memcon'15 (2015)

\bibitem{al2005dram}
Al-Ars, Z.: DRAM fault analysis and test generation. TU Delft (2005)

\bibitem{Aweke2016}
Aweke, Z.B., Yitbarek, S.F., Qiao, R., Das, R., Hicks, M., Oren, Y., Austin,
  T.: {ANVIL: Software-Based Protection Against Next-Generation Rowhammer
  Attacks}. In: {ASLPOS}'16 (2016)

\bibitem{bains2014row1}
Bains, K., Halbert, J.: Row hammer monitoring based on stored row hammer
  threshold value (Jun~5 2014), {US Patent App. 13/690,523}

\bibitem{bains2014row2}
Bains, K., Halbert, J., Mozak, C., Schoenborn, T., Greenfield, Z.: Row hammer
  refresh command (Jan~2 2014), {US Patent App. 13/539,415}

\bibitem{Barresi2015}
Barresi, A., Razavi, K., Payer, M., Gross, T.R.: {CAIN:} silently breaking
  {ASLR} in the cloud. In: {WOOT}'15 (2015)

\bibitem{2004-bernstein-cachetiming}
Bernstein, D.J.: {Cache-timing attacks on AES}. Tech. rep., Department of
  Mathematics, Statistics, and Computer Science, University of Illinois at
  Chicago (2005)

\bibitem{DBLP:conf/crypto/BihamS97}
Biham, E., Shamir, A.: Differential fault analysis of secret key cryptosystems.
  In: {CRYPTO} '97. LNCS, vol. 1294 (1997)

\bibitem{DBLP:conf/eurocrypt/BonehDL97}
Boneh, D., DeMillo, R.A., Lipton, R.J.: {On the Importance of Checking
  Cryptographic Protocols for Faults}. In: {EUROCRYPT}'97. LNCS, vol. 1233
  (1997)

\bibitem{Gruss2015esorics}
Gruss, D., Bidner, D., Mangard, S.: Practical memory deduplication attacks in
  sandboxed javascript. In: ESORICS'15 (2015)

\bibitem{FlushFlush}
Gruss, D., Maurice, C., Wagner, K., Mangard, S.: {Flush+Flush: A Fast and
  Stealthy Cache Attack}. In: DIMVA'16 (2016)

\bibitem{191010}
Gruss, D., Spreitzer, R., Mangard, S.: {Cache Template Attacks: Automating
  Attacks on Inclusive Last-Level Caches}. In: USENIX Security'15 (2015)

\bibitem{DBLP:conf/sp/GullaschBK11}
Gullasch, D., Bangerter, E., Krenn, S.: {Cache Games -- Bringing Access-Based
  Cache Attacks on {AES} to Practice}. In: S{\&}P'11 (2011)

\bibitem{Herath2015}
Herath, N., Fogh, A.: {These are Not Your Grand Daddys CPU Performance Counters
  - CPU Hardware Performance Counters for Security}. Black Hat  (2015)

\bibitem{Huang2012}
Huang, R.F., Yang, H.Y., Chao, M.C.T., Lin, S.C.: {Alternate hammering test for
  application-specific DRAMs and an industrial case study}. In: DAC'12 (2012)

\bibitem{Hund2013}
Hund, R., Willems, C., Holz, T.: {Practical Timing Side Channel Attacks against
  Kernel Space ASLR}. In: S{\&}P'13 (2013)

\bibitem{Inci2015iacr}
Inci, M.S., Gulmezoglu, B., Irazoqui, G., Eisenbarth, T., Sunar, B.:
  {Seriously, get off my cloud! Cross-VM RSA Key Recovery in a Public Cloud}.
  Cryptology ePrint Archive, Report 2015/898 pp. 1--15 (2015)

\bibitem{DBLP:conf/sp/IrazoquiES15}
Irazoqui, G., Eisenbarth, T., Sunar, B.: {{S\$A}: A Shared Cache Attack that
  Works Across Cores and Defies {VM} Sandboxing -- and its Application to
  {AES}}. In: S{\&}P'15 (2015)

\bibitem{Kim2014ISCA}
Kim, Y., Daly, R., Kim, J., Fallin, C., Lee, J.H., Lee, D., Wilkerson, C., Lai,
  K., Mutlu, O.: {Flipping bits in memory without accessing them: An
  experimental study of DRAM disturbance errors}. In: ISCA'14 (2014)

\bibitem{Lanteigne2016}
Lanteigne, M.: {How Rowhammer Could Be Used to Exploit Weakness Weaknesses in
  Computer Hardware} (March) (2016), \url{http://www.thirdio.com/rowhammer.pdf}

\bibitem{LippGSM15}
Lipp, M., Gruss, D., Spreitzer, R., Mangard, S.: Armageddon: Last-level cache
  attacks on mobile devices. CoRR  abs/1511.04897 (2015)

\bibitem{Liu2015}
Liu, F., Yarom, Y., Ge, Q., Heiser, G., Lee, R.B.: {Last-Level Cache
  Side-Channel Attacks are Practical}. In: S{\&}P'15 (2015)

\bibitem{Maurice2015RAID}
Maurice, C., Le~Scouarnec, N., Neumann, C., Heen, O., Francillon, A.: {Reverse
  Engineering Intel Last-Level Cache Complex Addressing Using Performance
  Counters}. In: RAID'15 (2015)

\bibitem{Maurice2015C5}
Maurice, C., Neumann, C., Heen, O., Francillon, A.: {C5: Cross-Cores Cache
  Covert Channel}. In: DIMVA'15 (2015)

\bibitem{MicronTN-46-09}
Micron: {Designing for 1Gb DDR SDRAM}.
  \url{https://www.micron.com/~/media/documents/products/technical-note/dram/tn4609.pdf}
  (2003)

\bibitem{Oren2015CCS}
Oren, Y., Kemerlis, V.P., Sethumadhavan, S., Keromytis, A.D.: {The Spy in the
  Sandbox: Practical Cache Attacks in JavaScript and their Implications}. In:
  CCS'15 (2015)

\bibitem{DBLP:conf/ctrsa/OsvikST06}
Osvik, D.A., Shamir, A., Tromer, E.: {Cache Attacks and Countermeasures: The
  Case of {AES}}. In: Topics in Cryptology -- {CT-RSA}. LNCS, vol. 3860 (2006)

\bibitem{Park2014}
Park, K., Baeg, S., Wen, S., Wong, R.: {Active-Precharge Hammering on a Row
  Induced Failure in DDR3 SDRAMs under 3x nm Technology}. In: IIRW'14 (2014)

\bibitem{Payer2016}
Payer, M.: {HexPADS}: a platform to detect ``stealth'' attacks. In: {ESSoS}'16
  (2016)

\bibitem{2005-percival-cache}
Percival, C.: {Cache missing for fun and profit}. In: Proceedings of BSDCan
  (2005)

\bibitem{Pessl2015}
Pessl, P., Gruss, D., Maurice, C., Mangard, S.: Reverse engineering intel
  {DRAM} addressing and exploitation. CoRR  abs/1511.08756 (2015)

\bibitem{Qureshi2007}
Qureshi, M.K., Jaleel, A., Patt, Y.N., Steely, S.C., Emer, J.: {Adaptive
  insertion policies for high performance caching}. ACM SIGARCH Computer
  Architecture News  35(2),  381 (2007)

\bibitem{Rahmati2015}
Rahmati, A., Hicks, M., Holcomb, D.E., Fu, K.: Probable cause: the
  deanonymizing effects of approximate {DRAM}. In: {ISCA}'15 (2015)

\bibitem{Seaborn2015DRAMmap}
Seaborn, M.: {How physical addresses map to rows and banks in DRAM}.
  \url{http://lackingrhoticity.blogspot.com/2015/05/how-physical-addresses-map-to-rows-and-banks.html}
  (May 2015), retrieved on July 20, 2015

\bibitem{SeabornBlackhat2015}
Seaborn, M., Dullien, T.: Exploiting the {DRAM} rowhammer bug to gain kernel
  privileges. In: Black Hat (2015)

\bibitem{2015W3C}
{W3C}: {High Resolution Time Level 2 - W3C Working Draft 21 July 2015}.
  \url{http://www.w3.org/TR/2015/WD-hr-time-2-20150721/#privacy-security} (Jul
  2015)

\bibitem{Wong2013}
Wong, H.: {Intel Ivy Bridge Cache Replacement Policy}.
  \url{http://blog.stuffedcow.net/2013/01/ivb-cache-replacement/}, retrieved on
  July 16, 2015

\bibitem{DBLP:conf/uss/YaromF14}
Yarom, Y., Falkner, K.: {{FLUSH+RELOAD:} {A} High Resolution, Low Noise, {L3}
  Cache Side-Channel Attack}. In: {USENIX} Security'14 (2014)

\bibitem{Yarom2015iacr}
Yarom, Y., Ge, Q., Liu, F., Lee, R.B., Heiser, G.: {Mapping the Intel
  Last-Level Cache}. Cryptology ePrint Archive, Report 2015/905 pp. 1--12
  (2015)

\end{thebibliography}

\end{document}